\documentclass[pdflatex,sn-mathphys,usenames,dvipsnames]{sn-jnl}

\usepackage[utf8]{inputenc} 
\usepackage[T1]{fontenc}    
\usepackage{hyperref}       
\usepackage{url}            
\usepackage{booktabs}       
\usepackage{amsfonts}       
\usepackage{nicefrac}       
\usepackage{microtype}      
\usepackage{lipsum}		
\usepackage{graphicx}
\usepackage{graphbox}
\usepackage[numbers]{natbib} 
\usepackage{doi}

\usepackage{listings}
\usepackage{subcaption}
\usepackage{amsmath}
\usepackage{wrapfig}
\usepackage{refcount}
\usepackage[htt]{hyphenat} 
\usepackage{multirow}
\usepackage{ulem}
\usepackage[flushleft]{threeparttable}
 \normalbaroutside 

\jyear{2021}%

\theoremstyle{thmstyleone}%
\theoremstyle{thmstyletwo}%

\theoremstyle{thmstylethree}%

\begin{document}

\title[SFILES 2.0: An extended text-based flowsheet representation]{SFILES 2.0: An extended text-based flowsheet representation}


\author[1]{\fnm{Gabriel} \sur{Vogel}}

\author[1]{\fnm{Edwin} \sur{Hirtreiter}}

\author[1]{\fnm{Lukas} \sur{Schulze Balhorn}}

\author*[1]{\fnm{Artur} \sur{Schweidtmann}}\email{a.schweidtmann@tudelft.nl}

\affil[1]{\orgdiv{University of Technology, Department of Chemical Engineering}, \orgname{TU Delft}, \orgaddress{Van der Maasweg 9 2629 HZ \city{Delft}, \country{The Netherlands}}}


\abstract{SFILES are a text-based notation for chemical process flowsheets. 
They were originally proposed by d'Anterroches \cite{d'Anterroches2006} who was inspired by the text-based SMILES notation for molecules. The text-based format has several advantages compared to flowsheet images regarding the storage format, computational accessibility, and eventually for data analysis and processing. However, the original SFILES version cannot describe essential flowsheet configurations unambiguously, such as the distinction between top and bottom products. Neither is it capable of describing the control structure required for the safe and reliable operation of chemical processes. 
Also, there is no publicly available software for decoding or encoding chemical process topologies to SFILES. 
We propose the SFILES~2.0 with a complete description of the extended notation and naming conventions. Additionally, we provide open-source software for the automated conversion between flowsheet graphs and SFILES~2.0 strings. 
This way, we hope to encourage researchers and engineers to publish their flowsheet topologies as SFILES~2.0 strings. The ultimate goal is to set the standards for creating a FAIR database of chemical process flowsheets, which would be of great value for future data analysis and processing.}

\keywords{Flowsheet graph, Process flow diagram, Artificial intelligence, FAIR data, STRING notation}



\maketitle

\section{Introduction}\label{sec1}

Chemical process flowsheets, also known as process flow diagrams~(PFDs) \cite{ISO2010,ISO2015}, are the current standard for depicting and communicating the topology of unit operations in chemical processes (see Figure~\ref{fig:Flowsheet_intro}). 
PFDs are used in industry and academia during conceptual process design and consequently there exists at least one PFD for every chemical process in the world. 
Besides process flow diagrams, Piping and Instrumentation Diagrams~(P\&IDs)~\cite{ISO2010,ISO2015} are a central representation class of chemical processes. They include additional information about instrumentation, valves, control structures, and piping~\cite{Towler2008}. Due to contained process-specific knowledge P\&IDs provide valuable details for a deep understanding of the chemical process. Therefore, P\&IDs are interdisciplinary employed at every stage of a chemical plant: from engineering and design, to hazard and operability studies (HAZOP), to operation and tracking changes during maintenance~\cite{Toghraei2019}.
Currently, PFDs and P\&IDs are usually drawn in computer programs and exported as images or PDF documents. 
Despite some recent efforts in Smart P\&IDs and open data exchange formats~\cite{Wiedau2019}, it seems that the information content of flowsheet diagrams in documents often remains inseparable from the medium, like hieroglyphs carved in stone. 
The main reason for this development is that PFDs and P\&IDs in the form of images or PDFs are widely utilized as an interdisciplinary communication tool for easily exchanging first process ideas, but also advanced plant designs between experts from different domains (e.g. process engineers, material scientists, management, etc.).
Also, proprietary process simulation software often does not facilitate interoperability and data exchange. 
However, the document-based communication of flowsheet information hinders the development of findable, accessible, interoperable, and reusable~(FAIR)~\cite{Wilkinson2016} data. 
This also has consequences for the use of advanced data analysis and data processing tools. 
Currently, some aspects of chemical process design can be tedious and repetitive, while FAIR process data could enable automated data processing. 
In our previous work, we also argue that the lack of structured data is a major hurdle for advances of artificial intelligence in chemical process engineering~\citep{Schweidtmann2021}. 

Chemical flowsheets can be represented as directed graphs~\citep{Zhang2018,Zheng2022}. The flowsheet graph (see Figure~\ref{fig:intro_pfd_graph}) consists of nodes that represent the unit operations and directed edges that represent the stream connections. Graphs are computationally accessible and further offer the possibility to store additional process information as node or edge attributes. However, using the graph as flowsheet representation usually requires knowledge of programming languages and graph libraries, both for the process designer and for engineers who want to reuse the flowsheet. 

Text-based representations are a promising alternative to graph representations for the communication of flowsheet information.
In 2006, d'Anterroches~\cite{d'Anterroches2006} proposed the Simplified Flowsheet Input-Line Entry-System~(SFILES) which is a text-based notation to represent flowsheet topologies. 
The SFILES is inspired by the Simplified Molecule Input-Line Entry-System~(SMILES)~\citep{Weininger1989} notation, which has become a standard storage and exchange format for molecules.
Using SFILES as flowsheet storage and exchange format brings several advantages compared to images and graphs. Standardization of the text-based representation is one advantage over flowsheet images that usually vary due to different drawing software. Furthermore, the text-based representation is an efficient exchange format that can be included in publications and directly used for data analysis and processing, which sets it apart from the graph representation. 

SFILES have already enabled the development of advanced data processing techniques on flowsheets. 
\cite{Tula2019,Tula2019a} used it to compare process flowsheets for a given synthesis problem. Their approach enabled them to find more sustainable process alternatives.
In other work, the SFILES notation was slightly modified and used for pattern recognition in chemical process flowsheets~\citep{Zhang2018,Zheng2022}. With the help of sequence alignment algorithms, the authors successfully identified common design patterns in chemical process flowsheets. Nevertheless, previous work does not include a complete description of the connectivity and the stream paths when dealing with unit operations with multiple in- and outlet streams, i.e., the distinction between top and bottom products or stream paths through multi-stream heat exchangers. Furthermore, the SFILES notation in previous work is limited to PFDs, neglecting important information contained in P\&IDs, such as control structures. To the best of our knowledge, there is also no publicly available software for the automated conversion between flowsheet graphs and SFILES~2.0 strings.

In this work, we propose the SFILES~2.0 and provide a comprehensive description of the extensions and modifications compared to previous work. Moreover, we suggest naming conventions to pave the way toward standardized SFILES strings.
The extensions in this paper include a set of rules for the flowsheet graph representation, specifying a new way to unambiguously represent multi-stream heat exchangers and unit operations with top and bottom in- and outlet streams in the flowsheet graph. Subsequently, we modified and extended the original SFILES notation rules, which allow an unambiguous string representation and enable a reversible conversion between a flowsheet graph and its corresponding SFILES~2.0 string. Eventually, it should be possible to describe flowsheet topologies of higher complexities while still encoding all necessary topological information in the SFILES~2.0 string. Additionally, we address the inclusion of control structures contained in P\&IDs in the flowsheet graph and SFILES~2.0 notation.
Moreover, we implemented a conversion algorithm in Python and made it openly accessible in a GitHub repository~\cite{Vogel2022} with illustrative examples, encouraging researchers to publish their future chemical process flowsheets with the corresponding SFILES~2.0 strings. This way, we hope to contribute to creating and continuously extending a machine-readable SFILES~2.0-based database of chemical process flowsheet topologies.

\section{Background}
\label{sec:background}

The following outlines previous work on the flowsheet graph representation and SFILES notation rules, which lays the foundation for our work. 

\subsection{Flowsheet graph representation}
\begin{figure}[h!]
\centering
\begin{subfigure}[b]{0.75\textwidth}
   \includegraphics[width=1\linewidth]{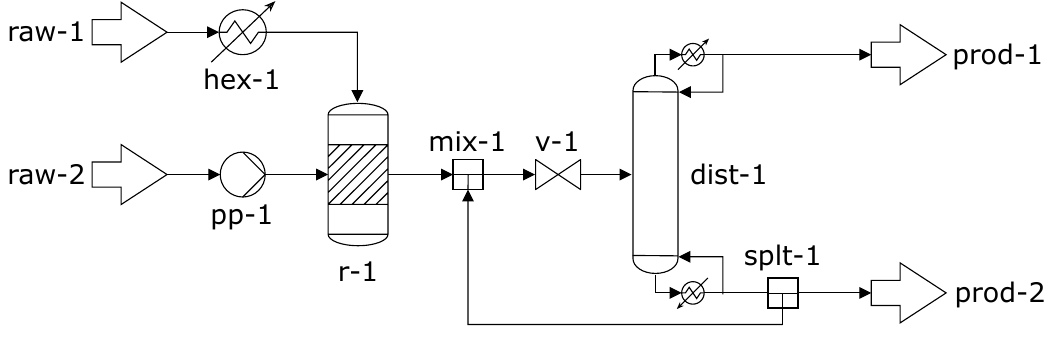}
   \caption{}
   \label{fig:Flowsheet_intro} 
\end{subfigure}

\begin{subfigure}[b]{0.85\textwidth}
   \includegraphics[width=1\linewidth]{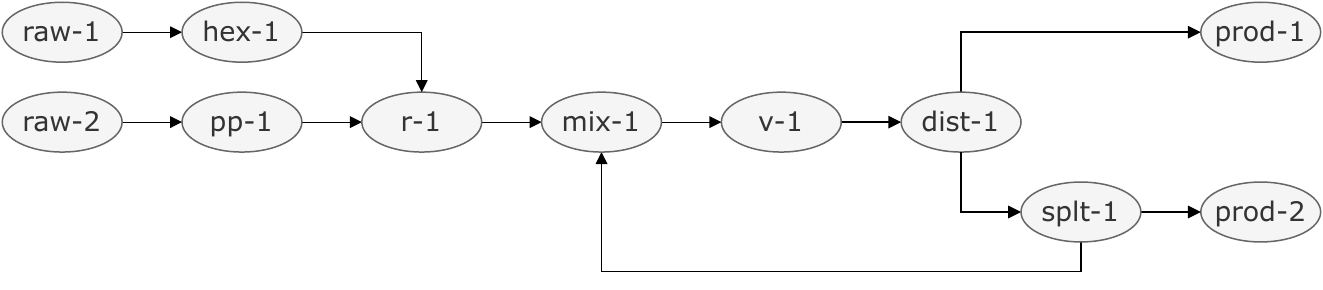}
   \caption{}
   \label{fig:intro_pfd_graph}
\end{subfigure}
\caption{(a) Simple chemical process flowsheet with branches and one recycle stream. (b) Graph representation of the flowsheet in (a)}

\end{figure}

A graph is a data structure that consists of nodes, also called vertices, and edges. Edges are connections between nodes and can be either directed or undirected, defining whether the graph is directed or undirected. 
The original description of the SFILES string~\citep{d'Anterroches2006} uses a directed flowsheet graph with process groups as nodes and the connections between these process groups as edges. The process groups can either represent one unit operation or a set of unit operations.
Herein, we focus on single unit operations in flowsheets, similar to the work in \cite{Zhang2018} defining unit operations as nodes and the connecting streams as edges. 
Figure~\ref{fig:Flowsheet_intro} shows an exemplary flowsheet with two inlet streams, a reactor, a distillation column~(reboiler and condenser included), a recycle of the bottom product, and two product streams. The used abbreviations are based on the standardized unit operation names in Table~\ref{tab:unit_operations}. When constructing the corresponding flowsheet graph in Figure~\ref{fig:intro_pfd_graph}, the nodes need to be numbered
to obtain a unique definition of nodes and their associated edges. We can distinguish the graph nodes using their in- and out-degree, whereby the in-degree is the number of edges directed towards a node, and the out-degree is the number of edges directed away from a node. Inlet nodes with the name \texttt{raw} always exhibit an in-degree=0, and outlet nodes with the name \texttt{prod} always have an out-degree=0. A node with an in-degree>1 means that graph branches are converging at that node~(in Figure~\ref{fig:intro_pfd_graph} \texttt{r-1, mix-1}), and a node with an out-degree>1 indicates a new branching at the considered node~(in Figure~\ref{fig:intro_pfd_graph} \texttt{dist-1, splt-1}). 

\subsection{Original SFILES notation}
\label{subsec:original_notation}
The original SFILES notation rules \citep{d'Anterroches2006} are outlined in this section using the flowsheet graph in Figure~\ref{fig:intro_pfd_graph}. 
Starting with the inlet node raw-1 the corresponding SFILES string of this flowsheet graph is
\begin{lstlisting}
(raw-1)(hex-1)(r-1)[<(pp-1)<(raw-2)](mix-1)<1(v-1)(dist-1)[(prod-1)](splt-1)1(prod-2).
\end{lstlisting}
Process groups, or in this example, abbreviations for the unit operations are noted in parenthesis. 
The SFILES string is read from left to right and two consecutive unit operations in parenthesis imply a connection, e.g., \texttt{(raw-1)(hex-1)} implies a connection from \texttt{(raw-1)} to \texttt{(hex-1)}. 
In the case of branching in the graph, e.g., after the distillation system (dist-1) in Figure~\ref{fig:Flowsheet_intro}, all except the last considered branch during the conversion from flowsheet graph to string (see Section~\ref{sec:graph_invariant}), are noted in square brackets. 
In the case of converging branches at a node with an in-degree>1, the original definition~\cite{d'Anterroches2006} uses square brackets and \texttt{<} for backward connections in the SFILES string. Converging branches always occur when the described chemical process comprises multiple input streams.
Consequently, the sequence \mbox{\texttt{(r-1)[<(pp-1)<(raw-2)]}} implies the connections from \texttt{(raw-2)} to \texttt{(pp-1)} and \texttt{(pp-1)} to \texttt{(r-1)}. 
The last important notation rule applies to recycle connections, such as the one from \texttt{(splt-1)} back to \texttt{(mix-1)}. 
Similar to cycles in molecules in the SMILES notation, a number \# is used to indicate the start of a recycle (here: (splt-1)1), and <\# is used to indicate the end of the directed recycle connection (here: (mix-1)<1). 
Given the flowsheet graph, the SFILES string generation consists of two steps~\citep{d'Anterroches2006}: 
\begin{enumerate}
    \item Calculation of a unique graph invariant.
    \item SFILES generation by traversing the graph with initial node selection and branching decisions based on the graph invariant.
\end{enumerate}
The graph invariant calculation is based on the flowsheet graph structure and is used to assign a unique rank to each node (see Section~\ref{sec:graph_invariant}). Based on the node ranks, an initial node for the graph traversal is chosen and branching decisions are made. This ensures the generation of a unique SFILES string.

The numbers in a SFILES string are adopted from the node names in the flowsheet graph but do not contain any essential process knowledge. For this reason, in previous work~\citep{Zhang2018, Zheng2022} for pattern recognition in flowsheets, the authors used a generalized version of the SFILES string without the unit operation numbers. Removing the numbering in the SFILES string of the example in Figure~\ref{fig:intro_pfd_graph} yields the generalized SFILES
\begin{lstlisting}
 (raw)(hex)(r)[<(pp)<(raw)](mix)<1(v)(dist)[(prod)](splt)1(prod).
\end{lstlisting}

\section{SFILES~2.0}
In this section, we describe our proposed modifications and extensions of the original SFILES notation. We call this modified version SFILES 2.0. Section~\ref{sec:extensions} clarifies minor modifications of the syntax and proposes extensions to unambiguously represent multi-stream heat exchangers and unit operations with complex connectivity, such as separation columns.
Thereafter, Section~\ref{subsec:control_structure_extensions} describes the notation details that are required to represent the control structure contained in P\&IDs. 
Additionally, we propose standardized naming conventions for commonly used unit operations in Section~\ref{subsec:unit-ops}. Finally, to enhance the usability for other researchers in the field, we created Tables \ref{tab:rules} and \ref{tab:rules_ctrl} in Section~\ref{sec:rules-summary} summarizing the SFILES~2.0 syntax and notation rules.
In the following, we use generalized SFILES as the standard notation (see Section~\ref{subsec:original_notation}). 

\subsection{Extension of notation}\label{sec:extensions}

\begin{figure}[h!]
	\centering
	\includegraphics[width=0.65\textwidth]{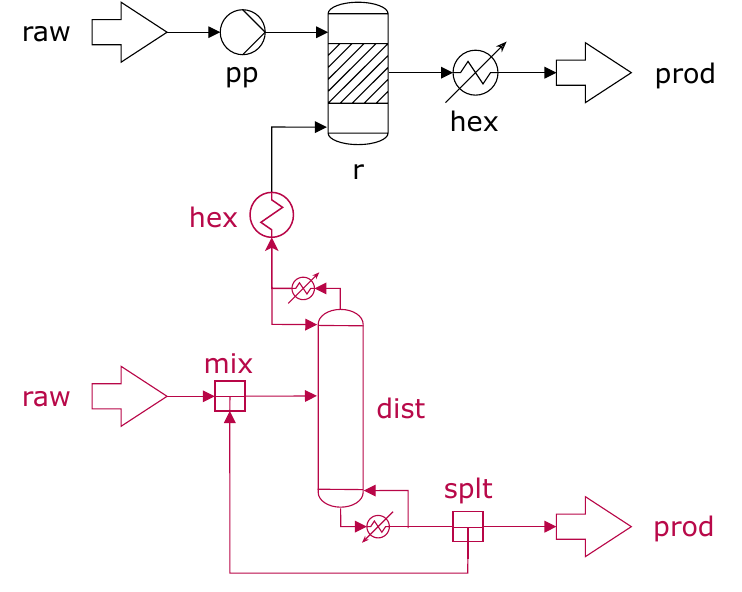}
	\caption{Flowsheet with multiple branchings.The branch in dark red colour is a converging branch in the SFILES~2.0 string consisting of a branching itself. It illustrates the necessity of our modification of adding a \& sign to encode how the branches are connected.}
	\label{fig:Converging_branch}
\end{figure}

\begin{figure}[h!]
	\centering
	\includegraphics[width=\textwidth]{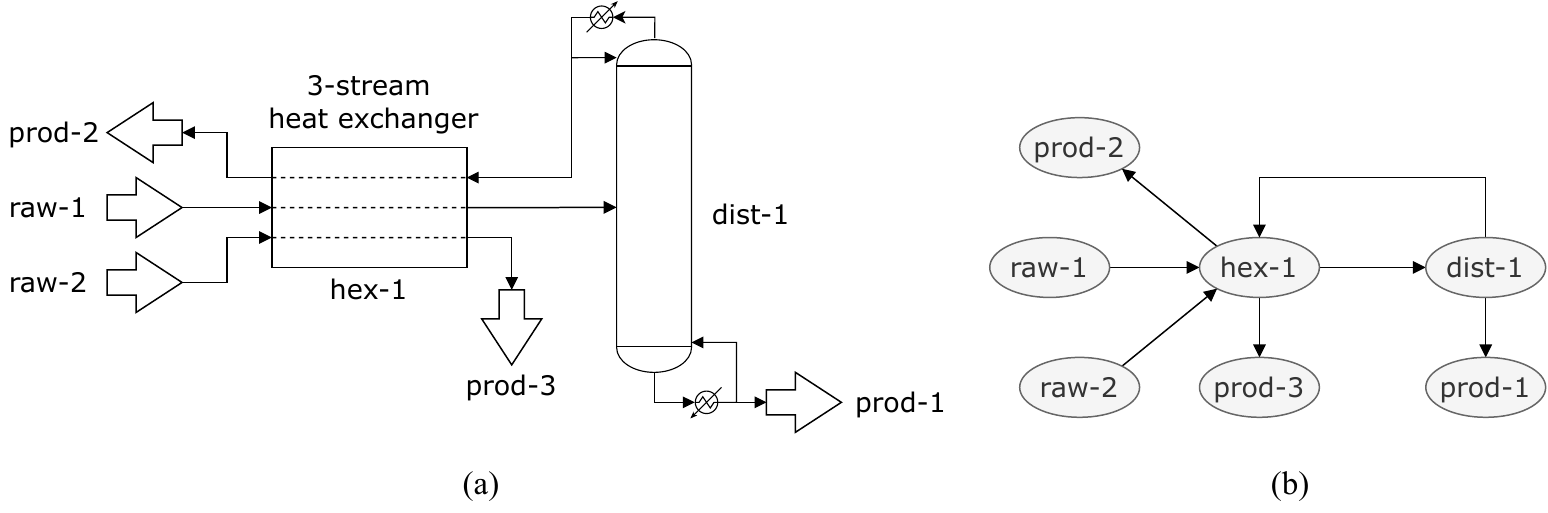}
	\caption{Flowsheet with complex connectivity characteristics. (a)~PFD, (b)~graph representation}
	\label{fig:connectivity}
\end{figure}

For complex chemical processes, the corresponding flowsheets can get quite large, containing a high number of unit operations and process branches. 
For a more robust notation of complex converging branches (multiple input streams), we suggest the following modification:
When reaching a node with an in-degree>1 during the graph traversal~(see Section~\ref{subsec:graph_traversal}), the original SFILES definition uses a backward notation containing < signs for converging branches. In the SFILES~2.0, we note converging branches surrounded by <\&| and |, whereby we insert an additional \&-sign next to the node that is connected to the considered node with an in-degree>1. 
Using this notation for the example in Figure~\ref{fig:intro_pfd_graph} yields the generalized SFILES
\begin{lstlisting}[mathescape=true]
(raw)(hex)(r)<&$|$(raw)(pp)&$|$(mix)<1(v)(dist)[(prod)](splt)1(prod)
\end{lstlisting}
It eliminates the backward notation containing < signs and, more importantly, enables a more robust notation of complex converging branches that consist of several branches themselves. An example that illustrates the necessity of this modification is shown in the flowsheet in Figure~\ref{fig:Converging_branch}. Using the conversion algorithm described in Section~\ref{subsec:graph_traversal} the SFILES~2.0 string for this flowsheet is:
\begin{lstlisting}[mathescape=True]
(raw)(pp)(r)@=<&$\textcolor{blue}{|}$=(raw)(mix)<1(dist)[(hex)=&=](splt)1(prod)$\textcolor{blue}{|}$@(hex)(prod)
\end{lstlisting}
The process branch that converges into the reactor is marked dark red in the SFILES~2.0 string and in the figure. According to the notation rules it is surrounded by \textcolor{blue}{\texttt{<\&|}} and \textcolor{blue}{\texttt{|}} (highlighted in dark blue). The additional \textcolor{blue}{\texttt{\&}} sign indicates which node of the dark red branch is connected to the reactor.
Furthermore, the following extensions in the SFILES~2.0 compared to previous work focus on how to describe the connectivity and the stream paths when dealing with unit operations with multiple in- and outlet streams. A common process characteristic that illustrates the importance of the connectivity information is heat integration, resulting in multi-stream heat exchangers. 
For instance, cryogenic processes such as air separation often comprise multi-stream heat exchangers. 
Other examples exhibiting complex connectivity are distillation columns with top and bottom products or even several inlet and outlet streams. 
The information on how the different streams are connected to the unit operations and are further processed is
essential and must be included in the SFILES~2.0 string to enable a reversible reconstruction of the flowsheet graph. 
The example process in Figure~\ref{fig:connectivity} consists of a 3-stream heat exchanger and a distillation column with top and bottom products. Essential information, in this case, is that the inlet \texttt{raw-1} is connected to the column via the heat exchanger and the top product is returned to the heat exchanger. 
Converting the flowsheet to a directed graph and to the generalized SFILES string without connectivity information yields
\begin{lstlisting}[mathescape=true]
(raw)(hex)<1<&$|$(raw)&$|$[(prod)][(prod)](dist)1(prod).
\end{lstlisting}
Using this SFILES string for the conversion back to a flowsheet graph would be ambiguous in terms of tracking the stream paths through the heat exchanger and the information of which separation product is heat-integrated with the heat exchanger.

\begin{figure}[h!]
\centering
\begin{subfigure}[b]{0.45\textwidth}
\centering
   \includegraphics[width=\textwidth]{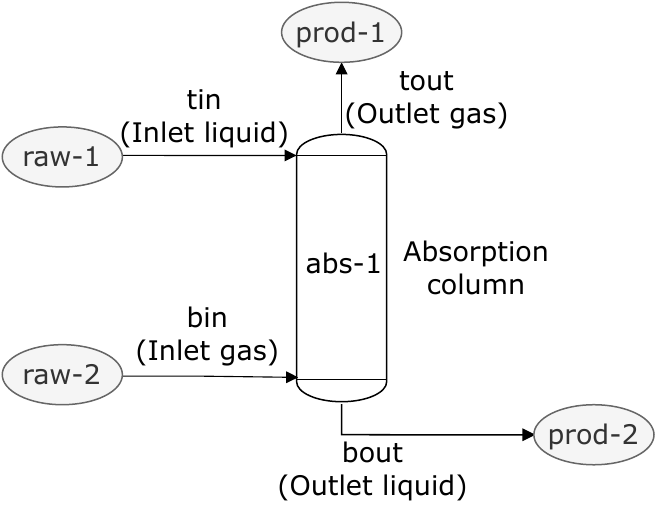}
   \caption{}
   \label{fig:absorption_pfd} 
\end{subfigure}
\hspace{1em}
\begin{subfigure}[b]{0.5\textwidth}
   \includegraphics[width=\textwidth]{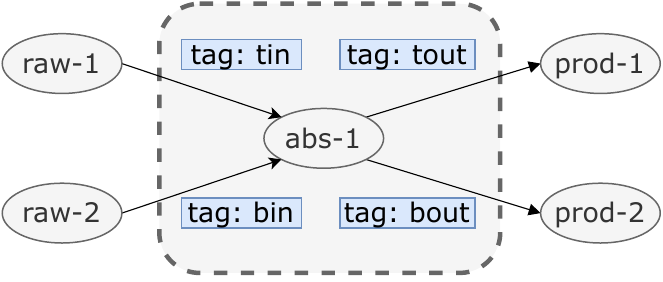}
   \caption{}
   \label{fig:absorption_graph}
\end{subfigure}
\caption{(a) Absorption column with two inlets and two outlets. (b) Flowsheet graph of (a) with connectivity stream tags}
\end{figure}

There are several possibilities to include the necessary information in the SFILES. Our strategy starts with modifying the flowsheet graph representation derived from the concept of State-Equipment Networks~(SEN)~\citep{Zhang2018, Yeomans1999} for the representation of the superstructure of chemical processes. As shown in Figure~\ref{fig:SEN_connectivity_graph}, we replaced the heat exchanger node with three single nodes that represent the accommodated streams in the heat exchanger equipment. Each node represents one heating or cooling task in that heat exchanger, meaning that the streams are not in direct contact but only transfer heat. We distinguish the node names in the graph by adding a \texttt{/\#}. Consequently, it is possible to have multiple separate mass trains resulting in multiple unconnected sub-graphs in the flowsheet graph. For instance, one sub-graph for the main process and one sub-graph for a refrigeration cycle. We will use the prefix \texttt{n|} in the SFILES string to indicate an independent mass train. In our example, one independent mass train is the connection from \texttt{raw-2} through \texttt{hex-1/3} to \texttt{prod-3}. In the numbered SFILES string, the node names of the heat exchangers contain the heat integration information. In the generalized SFILES string, we need to add this information after removing the numbers. The authors in~\cite{Zhang2018} used the recycle notation for heat integrated heat exchangers. However, the streams in heat exchangers do not mix, hence, formally this is not a recycle and we propose an alternative notation. Next to each heat exchanger node of the same heat exchanger equipment, we insert the same number \# in braces~(\texttt{\{\#\}}). In the case of heat exchangers~(heaters and coolers) without heat integration~(node has in-degree=1 and out-degree=1), we do not encode this information. Including the new rules for multi-stream heat exchangers, the following string results: 
\begin{lstlisting}[mathescape=true]
(raw)(hex){1}(dist)[(prod)](hex){1}(prod)n$|$(raw)(hex){1}(prod).
\end{lstlisting}
\begin{figure}
	\centering
	\includegraphics[width=0.55\textwidth]{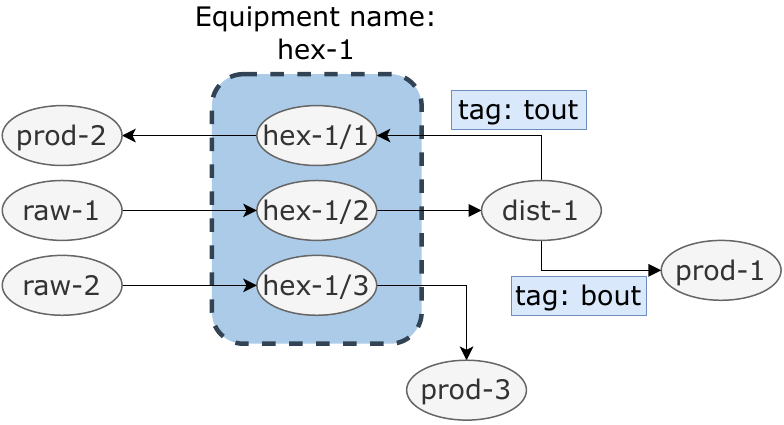}
	\caption{Flowsheet graph with modified node structure of heat exchanger and connectivity attributes for distillation column}
	\label{fig:SEN_connectivity_graph}
\end{figure}
We also need to encode additional information for other unit operations, such as distillation columns with at least one top and one bottom product as outlet streams. We use tags in the SFILES string to indicate the top product
branch and the bottom product branch. The difference between, for example, a column and a splitter is that the branched streams after a splitter have the same properties, whereas this, in general, does not hold for separation units. As a result, it is crucial information of the flowsheet topology which process branch results from which separation product. 
We will use braces to encode that additional connectivity information in the following manner. Given a graph edge from node u to node v with a stream tag x, the connection will be noted as 1. in case of a normal connection, 2. in case of branching, 3. in case of a recycle, and 4. in case of a converging branch.
\begin{enumerate}
    \item \texttt{(u)\{x\}(v)}
    \item \texttt{(u)[\{x\}(v)]}\quad or\quad \texttt{(u)[...]\{x\}(v)} 
    \item \texttt{(v)<1...(u)\{x\}1}
    \item \texttt{(v)<\&|...(u)\{x\}\&|}
\end{enumerate}
The stream tags must be saved as edge attributes in the flowsheet graph, e.g., the top and bottom outlet stream tags in Figure~\ref{fig:SEN_connectivity_graph}. Combining the rules related to multi-stream heat exchangers and stream tags, the final SFILES~2.0 string results in
\begin{lstlisting}[mathescape=true]
(raw)(hex){1}(dist)[{bout}(prod)]{tout}(hex){1}(prod)n$|$(raw)(hex){1}(prod).
\end{lstlisting}
The SFILES~2.0 string now enables the reconstruction of the flowsheet graph without loss of information and ultimately the reproduction of the PFD in Figure~\ref{fig:connectivity}.
The stream tags can also be applied to other unit operations such as absorption or extraction columns. Figure~\ref{fig:absorption_pfd} shows an absorption column with two inlet and two outlet connections. The necessary topological information is contained in the tags \texttt{\{bin\}}, \texttt{\{tin\}}, \texttt{\{tout\}}, and \texttt{\{bout\}}, which are stored as edge attributes in the flowsheet graph in Figure~\ref{fig:absorption_graph}. The SFILES~2.0 string for this hypothetical process is
\begin{lstlisting}[mathescape=true]
(raw){bin}(abs)<&$|$(raw){tin}&$|$[{tout}(prod)]{bout}(prod).
\end{lstlisting}
The same can be applied to all other units operation nodes where the connectivity information is considered essential for the flowsheet topology. Table~\ref{tab:tags} lists the defined tags. 
\begin{table}[h!]
	\caption{Set of stream and control tags used in SFILES~2.0. A complete list of possible control tags according to DIN~EN~62424 can be found in \cite{Winter2015}.}
	\centering
	\begin{tabular}{ll}
		\toprule
		& Stream tag in flowsheet graph\\ 
		Connectivity information&and SFILES~2.0 \\
		\midrule
		Bottom inlet&bin\\
		Top inlet&tin\\
		Bottom outlet&bout\\
		Top outlet&tout\\
        Multi-stream heat exchanger identifier& \# (unique number per heat exchanger)\\
		\midrule
        & Control tag in flowsheet graph\\ 
		Control structure information&and SFILES~2.0 \\
        \midrule
        Flow control& FC\\
        Level control&LC\\
        Pressure control&PC\\
        Temperature control&TC\\
        \bottomrule
	\end{tabular}
	\label{tab:tags}
\end{table}
\subsection{Description of control structures}\label{subsec:control_structure_extensions}
To extend the described text-based notation of PFDs to P\&IDs, a representation of the control structure is required. There are three important cases to consider for this: (i) A sensor on a stream controlling a unit operation, (ii) a sensor on a unit operation controlling another unit operation, and (iii) cascading sensors. We introduce the SFILES~2.0 notation for control structure by three illustrative examples in Figure~\ref{fig:SFILESctrl}. The first example (i) in Figure~\ref{fig:SFILESctrl}~(a) consists of a sensor measuring the flow rate of a stream and controlling the subsequent valve with this information. Since material streams are implicitly represented in the SFILES~2.0 notation, the measurement of stream information is included by adding the control unit (abbrev.~C) between the two unit operations (here \texttt{raw} and \texttt{prod}), where the state of a stream is required. The control unit is stored as a node like a unit operation. The type of the control unit, which is indicated in the P\&ID with a letter code (acc. to DIN~EN~62424)~\cite{Winter2015}, is stored in braces next to the node (here \{FC\} for flow control). Similar to material recycle connections, we represent signal connections to previous unit operations with <\_\# and \_\#. The underscore is used to easily distinguish material recycles and signal connections. Furthermore, we use upper case letters for control elements to illustrate the difference to unit operations. These notation rules result in the following generalized SFILES~2.0 for Figure~\ref{fig:SFILESctrl}~(a):
\begin{lstlisting}[mathescape=true]
(raw)(C){FC}_1(v)<_1(prod).
\end{lstlisting}
The second example (ii) in Figure~\ref{fig:SFILESctrl}~(b) shows a tank whose level is controlled. The direct connection of the instrument to the unit operation is represented as branching at the corresponding node. In the same way as for the first example, the letter code of the control unit is stored as a tag and the instrument is connected to the valve using the signal connection terminology: 
\begin{lstlisting}[mathescape=true]
(raw)(tank)[(C){LC}_1](v)<_1(prod).
\end{lstlisting}
The third example (iii) in Figure~\ref{fig:SFILESctrl}~(c) of a control cascade illustrates a combination of the first two cases. The level of the tank is transmitted to a flow controller, which regulates a subsequent valve. The flow transmitter is represented as a branching node at the corresponding unit operation and the flow controller is placed between the tank and valve since its task is to measure the flow rate between the tank and the valve. The connection of the two instruments and the valve is represented with the signal connection notation. Tags store again the letter code of the control units. This results in the following generalized SFILES~2.0 string for Figure~\ref{fig:SFILESctrl}~(c):
\begin{lstlisting}[mathescape=true]
(raw)(tank)[(C){LT}_1](C){FC}_2<_1(v)<_2(prod).
\end{lstlisting}
\begin{figure}[h!]
	\centering
	\includegraphics[width=\textwidth]{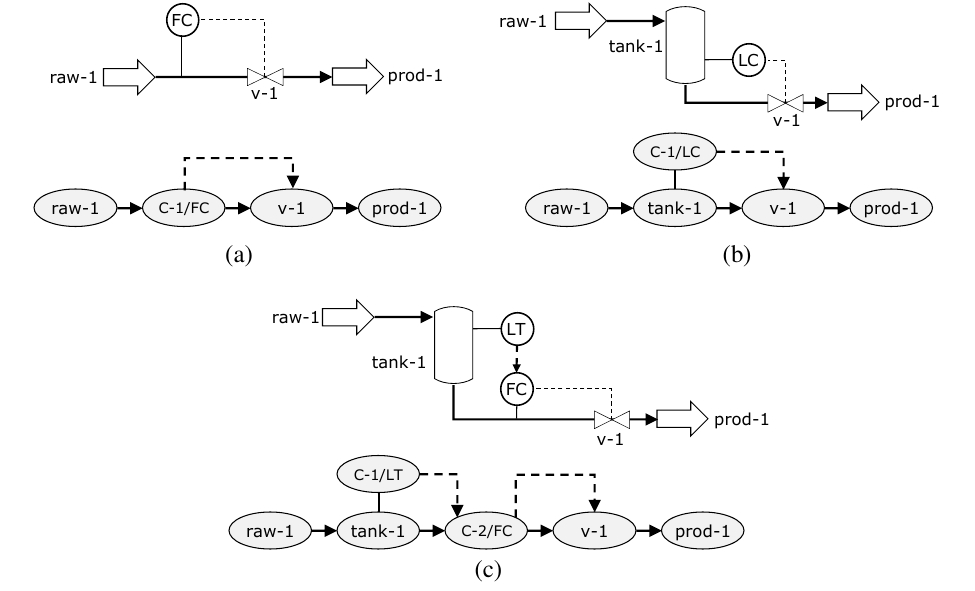}
	\caption{PFD and flowsheet graph of simple control loops. (a) Flow control of material stream, (b) Level control of tank, (c) Level control of tank with control cascade}
	\label{fig:SFILESctrl}
\end{figure}
\subsection{Unit operations}
\label{subsec:unit-ops}
This section provides an overview of unit operations in chemical process flowsheets and the abbreviations used in the SFILES~2.0. The selection of unit operations in Table~\ref{tab:unit_operations} represents commonly used unit operations and is based on the ontology OntoCAPE~\citep{Morbach2009}. Some of the terms in Table~\ref{tab:unit_operations} are a refined classification of the OntoCAPE ontology, which we performed to include more specific unit operation categories. With increasing access to more flowsheet data, the list of unit operations might need further extension or refinement.  
The naming conventions, i.e., the abbreviations, should also be followed in the flowsheet graph construction when using the provided code for the conversion from a flowsheet graph to its corresponding SFILES~2.0 string. 
\begin{table}[h!]
\centering
\begin{threeparttable}
	\caption{Unit operations and abbreviations in SFILES~2.0 based on OntoCAPE ontology~\citep{Morbach2009}}
	\centering
	\begin{tabular}{p{2.8cm}p{1.5cm}p{3.5cm}p{0.9cm}p{0.9cm}}
		\toprule
		&&&Typical&Typical\\
		Unit operation& Abbreviation&OntoCAPE term&in-degree&out-degree\\
		\midrule
		Absorption&abs&AbsorptionColumn&2&2\\
        Blower&blwr&Blower$^\mathrm{*}$ &1&1\\
        Centrifugation&centr&CentrifugationUnit&1&2\\
        Compressor&comp&Compressor$^\mathrm{*}$ &1&1\\
        Condenser\\~(incl. splitting)&cond&Condenser&1&2\\
        Control unit&C&Control&$\geq$1&$\geq$0\\
        Cyclone&cycl&Cyclone&1&2\\
        Distillation (incl. reboiler and condenser)&dist&DistillationSystem&$\geq$1&$\geq$2\\
        Electrical gas cleaning&egclean&ElectricalGasCleaningUnit&1&2\\
        Expander&expand&Expander$^\mathrm{*}$ &1&1\\
        Extraction&extr&ExtractionUnit&2&2\\
        Flash&flash&FlashUnit&1&$\geq$2\\
        Gas filtration&gfil&GasFilter&1&2\\
        Hydrocyclone&hcycl&Hydrocyclone&1&2\\
        Heat exchanger&hex&HeatExchanger&$\geq$1&$\geq$1\\
        Liquid filtration&lfil&LiquidFilter&1&2\\
        Mixing&mix&MixingUnit&$\geq$1&1\\
        Orifice plate&orif&OrificePlate$^\mathrm{*}$&1&1\\
        Pipe&pipe&Pipe$^\mathrm{*}$ &1&1\\
        Pump&pp&Pump$^\mathrm{*}$ &1&1\\
        Product stream&prod&OutputProduct&1&0\\
        Reactor&r&ChemicalReactor&$\geq$1&$\geq$1\\
        Raw material&raw&RawMaterial&0&1\\
        Reboiler\\~(incl. splitting)&reb&Reboiler&1&2\\
        Rectification (incl. reboiler and condenser)&rect&RectificationSystem&$\geq$1&$\geq$2\\
        Scrubbing&scrub&Scrubber&2&2\\
        Separation (no further sub-specification)&sep&SeparationUnit&$\geq$1&$\geq$2\\
        Splitting&splt&SplittingUnit&1&$\geq$2\\
        Stripping&strip&StrippingSystem$^\mathrm{*}$ &2&2\\
        Storage&tank&StorageUnit$^\mathrm{*}$&$\geq$0&$\geq$1\\
        Valve&v&Valve$^\mathrm{*}$&1&1\\
        Unknown unit&X&-&-&-\\
		\bottomrule
	\end{tabular}
	\begin{tablenotes}
      \item $^\mathrm{*}$\small This term is part of our extension of the OntoCAPE ontology and not in the original OntoCAPE documentation
    \end{tablenotes}
	\label{tab:unit_operations}
\end{threeparttable}
\end{table}

\subsection{Limitations of SFILES~2.0}
Nevertheless, there remain limitations of the SFILES~2.0 notation in the case of very complex process topologies. In the set of standardized stream tags for separation columns, we only consider top and bottom in- and outlets. The latter certainly covers the most common arrangements of unit operations in processes. Still, more complex examples such as the air separation process can contain columns with far more than two in- and outlets, respectively. For such complex unit operations, the current SFILES~2.0 notation rules do not suffice to ensure a reversible conversion between the SFILES string and the flowsheet in terms of the order of in- and outlets. At this point, we would like to mention that all types of flowsheets can be converted to an SFILES string. However, with a possible loss of information due to missing tags and, therefore, no fully reversible conversion back to the actual flowsheet. Theoretically, it would be possible to extend the notation to encode more complex information, e.g., by changing the stream tags to positions relative to the height of columns~(between 0 and 1, e.g., \texttt{\{1.0\_out\}} for the top outlet). Another approach could be to further divide equipment into several nodes, similarly to the SEN-based method for multi-stream heat exchangers. The braces notation could optionally also store flowsheet information beyond the topology in the SFILES~2.0. For instance, additional stream-related process information like the pressure, temperature, or components can be stored as edge attributes.

Additionally, information describing a unit operation, such as the geometrical dimensions or operating conditions are currently not stored in the SFILES~2.0 string. When desired, it could be stored as node attributes in the flowsheet graph and included in braces within the parentheses notation for unit operations. However, in this context, it must be pointed out that this information results in continuous variables which are not essential for describing the topology of flowsheets.

Furthermore, a more detailed description of the control structure, e.g., whether the instrument is a field-mounted or shared display device, is currently not provided.

\subsection{Summary of SFILES~2.0 rules}
\label{sec:rules-summary}
This subsection provides a summary of the SFILES~2.0 rules, i.e., Table~\ref{tab:rules} summarizes the general rules, whereas Table~\ref{tab:rules_ctrl} shows the notation rules specifically defined for P\&IDs.
\begin{table}[h!]
	\caption{Summary of SFILES~2.0 rules: PFD and P\&ID related}
	\centering
	\begin{tabular}{p{0.25\textwidth} c p{0.35\textwidth}}
		\toprule
		SFILES~2.0 sequence & Simplified graph & Rule explanation\\ 
		\midrule
        (u)(v) & \raisebox{-\totalheight}{\includegraphics[scale=0.7]{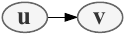}}& Two subsequent unit operations in parenthesis (u) and (v) imply a directed connection from (u) to (v) \\ \\
        (u)[(v)](w) & \raisebox{-\totalheight}{\includegraphics[scale=0.7]{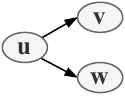}} & Square brackets indicate a branching. The unit operation (u) has two outlet connections to (v) and (w). Every branch except the last one (here, (w) is not in brackets) is noted in brackets. \\ \\
        (u)(v)<1(w)(x)1 & \raisebox{-\totalheight}{\includegraphics[scale=0.7]{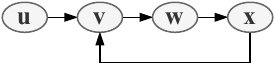}} & The symbols <\# and \# (\# is a number) indicate a recycle connection. In this case the recycle connection is from (x) to (v). \\ \\
        (w)(x)(v)<\&|(u)\&| & \raisebox{-\totalheight}{\includegraphics[scale=0.7]{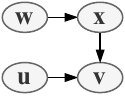}} & Incoming new branches are denoted between <\&| and \&|. In this example the incoming branch at (v) starts from a new inlet node (w); the \& sign indicates the connection from (x) to (v). \\ \\
        (u)(v)n|(w)(x) & \raisebox{-\totalheight}{\includegraphics[scale=0.7]{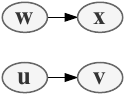}} & The n| indicates a new separate process (w)(x) with no material stream connection to the previous process (u)(v). \\ \\
        (u)(hex)\{1\}(v) n|(w)(hex)\{1\}(x) & \raisebox{-\totalheight}{\includegraphics[scale=0.7]{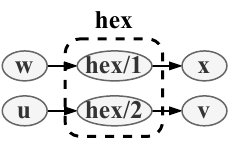}} & The notation {\#} after a heat exchanger unit is used for multi-stream heat exchangers to identify which streams share a specific heat exchanger (heat integration). \\ \\
        (u)[\{tout\}(v)]\{bout\}(w) & \raisebox{-\totalheight}{\includegraphics[scale=0.7]{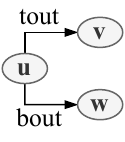}} & Stream tags are noted in braces, e.g. \{tout\} for top outlet, which is used to store information for unit operations with multiple in-/outlet streams, where the position of in-/outlet is important for the process description (e.g. distillation, absorption, ...).  \\
        \bottomrule
	\end{tabular}
	\label{tab:rules}
\end{table}

\begin{table}[h!]
	\caption{Summary of SFILES~2.0 rules: only P\&ID related}
	\centering
	\begin{tabular}{p{0.25\textwidth}c p{0.35\textwidth}}
		\toprule
		SFILES~2.0 sequence & Simplified graph& Rule explanation \\ 
		\midrule
        (u)(C)\{X\}(v) & \includegraphics[align=t,scale=0.7]{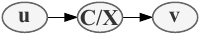} & Control units C for measuring the state of a stream, e.g. a flow indicator (FI), are noted between the two unit operations u and v where the measurement takes place. The letter code X is noted in braces directly after the control unit C, which is noted in parenthesis. \\ \\
        (u)[(C)\{X\}](v) & \includegraphics[align=t,scale=0.7]{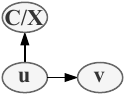} & The control unit C with the letter code X noted in square brackets indicate that the control unit, e.g. a level control (LC), is directly connected to a unit operation (here, u).  \\ \\
        (u)(C)\{X\}\_1(v)<\_1 & \includegraphics[align=t, scale=0.7]{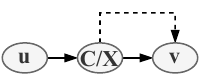}& Signal connections between control units or control units and unit operations are indicated with <\_\# and \_\#. The smaller than sign defines the direction: here, the control unit C with the letter code X is connected to the unit operation v. \\
        \bottomrule
	\end{tabular}
	\label{tab:rules_ctrl}
\end{table}

\section{SFILES~2.0 generation algorithm} \label{sec:generation}
This section describes the conversion algorithm between flowsheet graphs and SFILES~2.0 strings. 
Our implementation consists of the conversion algorithm from flowsheet graphs to SFILES~2.0 strings as well as the algorithm for the conversion of SFILES~2.0 strings to the corresponding flowsheet graphs and is publicly available in a GitHub repository~\cite{Vogel2022}.
Similar to the original SFILES notation algorithm (see Section~\ref{subsec:original_notation}), the two major steps for the SFILES~2.0 string generation are the determination of the graph invariant (Section~\ref{sec:graph_invariant}) and the graph traversal (Section~\ref{subsec:graph_traversal}). If a control structure is present in the flowsheet graph the nodes of the control units are treated as unit operation nodes. Only the signal connections (dashed line in P\&IDs) are removed before determining the graph invariant and the graph traversal, to ensure complete interoperability between SFILES~2.0 generated from P\&IDs and PFDs. The signal connections are added afterward using the notation mentioned in Section~\ref{subsec:control_structure_extensions}.

\subsection{Determination of graph invariant}
\label{sec:graph_invariant}
\begin{figure}[h!]
\centering
\includegraphics[width=\textwidth]{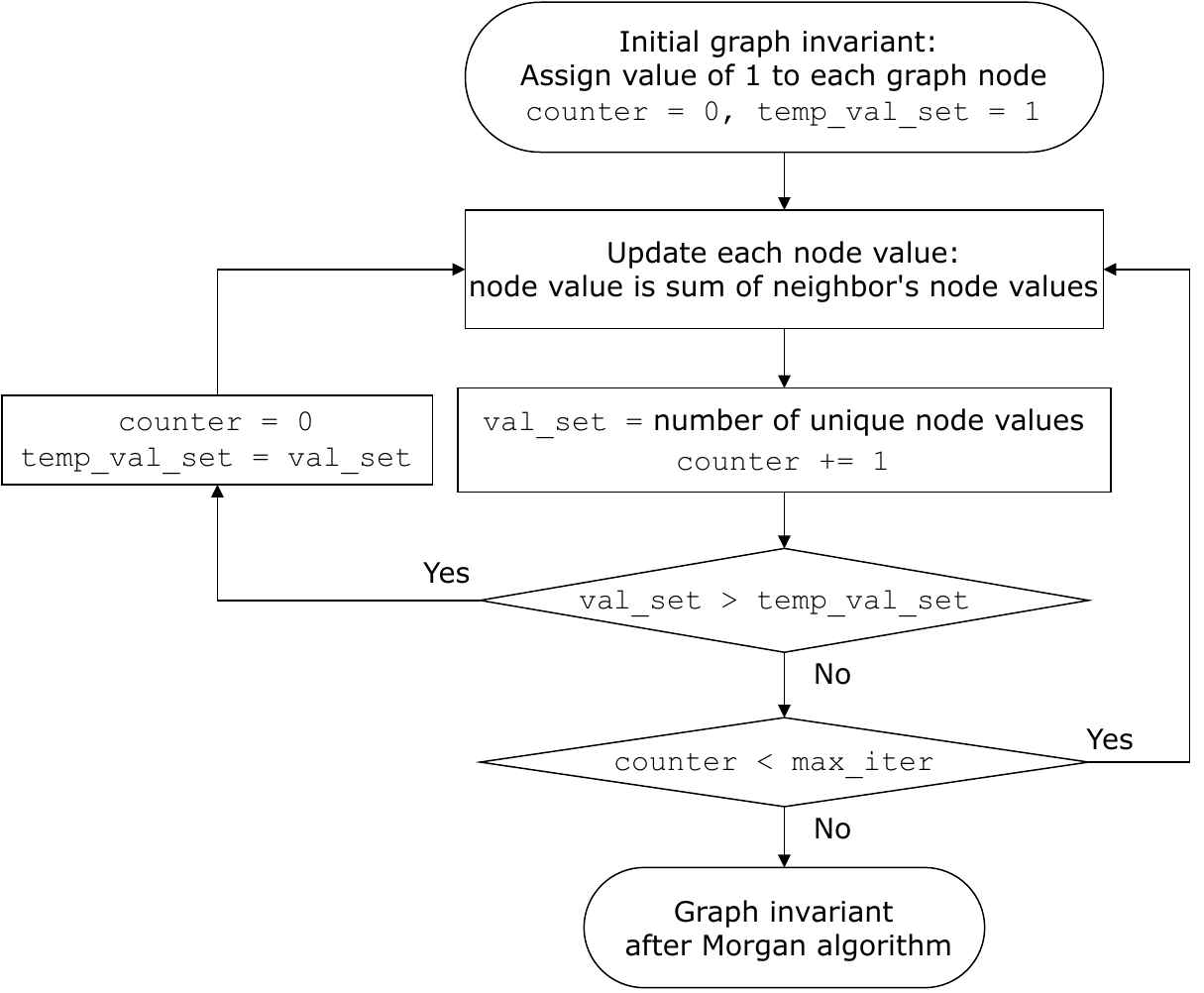}
\caption{Morgan algorithm for graph invariant determination}\label{fig:algorithm}
\end{figure}

The graph invariant aims to yield a unique rank for each node. The determination of this graph invariant is also known as graph canonization. The first step in our implementation is based on the Morgan algorithm~\citep{Morgan1965}, similarly to the description in~\cite{Zhang2018}. As illustrated in Figure~\ref{fig:algorithm}, the initialization starts with assigning all nodes a corresponding node value of 1. 
Next, each node value is updated with the sum of all neighbor's node values. After the first update, the node values equal their connectivity in the graph. This step aims to increase the variable \texttt{val\_set} which is defined as the number of unique node values in the graph. The procedure is repeated until \texttt{val\_set} does not increase for \texttt{max\_iter} iterations. Finally, the nodes are ranked based on their values. In case there are multiple sub-graphs, as described in Section~\ref{sec:extensions}, the graph invariant is determined for both graphs separately. The sub-graph with fewer nodes will be assigned a lower priority and noted last in the SFILES~2.0 string. 
The Morgan algorithm does not yield unique ranks in all cases. Especially in the case of symmetric graphs, there are often multiple nodes with the same value. However, the SFILES~2.0 string generation algorithm requires all nodes to have a unique rank. For this reason, we introduce a rule-based approach for breaking the ties of equally ranked nodes. We use the following procedure to break the ties. 
\begin{enumerate}
    \item Rank~(small is higher priority): Control node < Outlet node < Inlet node < Other nodes
    \item Rank according to the number of successors\footnote{The length of the depth first search tree of the node in the graph is used.} in the graph
    \begin{enumerate}
    \item Outlet/Control nodes: does not apply
    \item Inlet nodes: the higher the number of successors the lower the rank
    \item Other nodes: the lower the number of successors the lower the rank
    \end{enumerate}
    \item String comparison (smaller rank for earlier appearance in alphabet) of equally ranked node names~(unit operation abbreviations) and associated edges
    \item Ranking by graph node~(unit) numbering 
\end{enumerate}
In steps 1-3, we only use the generalized SFILES because the SFILES string should only be dependent on the intrinsic graph structure but not the numbering of the unit operations. Step 2 is subdivided into inlet and other nodes to improve the readability of the resulting SFILES string. Nodes still tied after step 3 can be exchanged arbitrarily without a resulting change in the generalized SFILES string. Therefore, in step 4, the nodes are ranked by their unique node names with unit numbering. Table~\ref{tab:node_ranking} shows the node ranking for the example in Figure~\ref{fig:intro_pfd_graph}. 
\begin{table}[h!]
	\caption{Node ranks for flowsheet graph in Figure~\ref{fig:intro_pfd_graph}}
	\centering
	\begin{tabular}{llllll}
		\toprule
		 \texttt{raw-1} &\texttt{raw-2}&\texttt{prod-1}&\texttt{prod-2}&\texttt{hex-1}&\texttt{pp-1} \\
		1&2&3&4&5&6\\
		\midrule
		\texttt{v-1}&\texttt{dist-1}&\texttt{r-1}&\texttt{splt-1}&\texttt{mix-1}& \\
		7&8&9&10&11&\\
		\bottomrule
	\end{tabular}
	\label{tab:node_ranking}
\end{table}
\subsection{Graph traversal} \label{subsec:graph_traversal}
The SFILES string results from traversing the graph after determining its invariant. We will use the depth-first search~(DFS) algorithm to traverse the flowsheet graph and write the SFILES string. 
Starting from an initial inlet node, the DFS algorithm explores the graph branches sequentially as far as possible~(until reaching an outlet node or previously visited node) before backtracking to the last branching point. 
Both the initial node selection as well as the branching decisions are made based on the node ranking, i.e., nodes with lower ranks are selected first. 
In the case of multiple inlet nodes or sub-graphs, one DFS traversal does not visit all nodes. To mitigate this problem a virtual node is inserted to which all initial nodes (in-degree=0) are connected. Since cycle processes do not exhibit a distinct initial node, the node with the lowest rank, which is not an outlet node (out-degree=0), is selected and connected to the virtual node. After ensuring that every node present in the flowsheet is linked to the virtual node, one graph traversal starting from the virtual node is sufficient.

Using the example in Figure~\ref{fig:intro_pfd_graph}, we will explain how the DFS algorithm and the SFILES string generation work. 
According to Figure~\ref{fig:intro_pfd_graph}, the nodes \texttt{raw-1} and \texttt{raw-2} with an in-degree=0 are connected to the virtual node and the graph traversal is started from there. Since \texttt{raw-1}, according to Table~\ref{tab:node_ranking}, has the lowest rank, the DFS visits this inlet node first.
The successor nodes, in specific \texttt{hex-1, r-1, mix-1, v-1, dist-1}, are visited one after another and noted in parentheses. 
After \texttt{dist-1} the top branch continues with \texttt{prod-1} (rank 3) and thereafter the bottom branch with \texttt{(splt-1)} (rank 10). Thus, the top branch starting with \texttt{prod-1} is visited first. The bottom branch leads to the mixer and after the second product \texttt{prod-2}, the first graph traversal ends. The resulting generalized SFILES~2.0 string is:
\begin{lstlisting}[mathescape=true]
(raw)(hex)(r)(mix)<1(v)(dist)[{tout}(prod)]{bout}(splt)1(prod).
\end{lstlisting}
The next node for the second graph traversal from the virtual node is \texttt{raw-2}. The branch converges in the reactor node \texttt{r-1} and the final generalized SFILES string is
\begin{lstlisting}[mathescape=true]
(raw)(hex)(r)<&$|$(raw)(pp)&$|$(mix)<1(v)(dist)[{tout}(prod)]{bout}(splt)1(prod)
\end{lstlisting}

\begin{figure}[h!]
	\centering
	\includegraphics[width=0.6\textwidth]{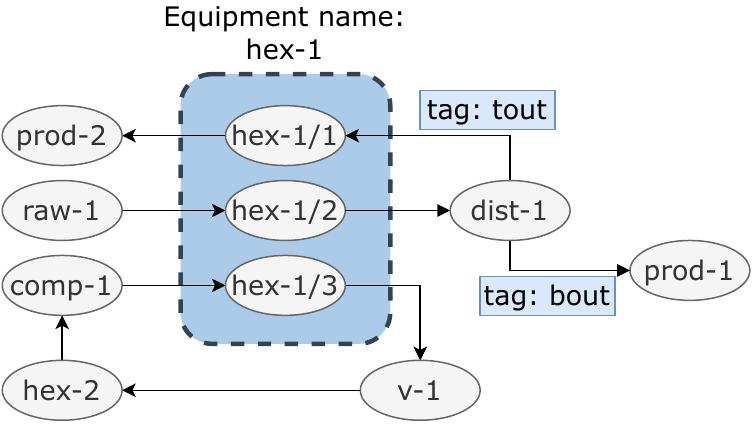}
	\caption{PFD graph with refrigeration cycle as sub-graph}
	\label{fig:SEN_connectivity_graph_2}
\end{figure}

Cycle processes are a special case of flowsheet topologies with no inlet nodes~(in\_degree=0). 
The cycle process can be either the complete flowsheet graph or a sub-graph, such as a refrigeration cycle. Assuming a refrigeration cycle instead of the stream from \texttt{raw-2} to \texttt{prod-3} in the example in Figure~\ref{fig:SEN_connectivity_graph} yields the modified graph in Figure~\ref{fig:SEN_connectivity_graph_2}. The graph traversal starting from the virtual node first explores the sub-graph containing the distillation system and results in
\begin{lstlisting}[mathescape=true]
(raw)(hex){1}(dist)[{bout}(prod)]{tout}(hex){1}(prod).
\end{lstlisting}
Since the nodes of the refrigeration cycle are still not visited, we need another DFS in this sub-graph. Because there is no inlet node in the refrigeration cycle, the node with the lowest rank which is not an outlet node (out-degree=0), in this case, \texttt{hex-1/3}, is connected to the virtual node and selected as the initial node. The final SFILES~2.0 string is
\begin{lstlisting}[mathescape=true]
(raw)(hex){1}(dist)[{bout}(prod)]{tout}(hex){1}(prod)n$|$(hex)<1(comp)(hex){1}(v)1.
\end{lstlisting}

\subsection{Conversion from SFILES~2.0 string to flowsheet graph}
The conversion of the SFILES~2.0 string back to a flowsheet graph is done by traversing the string and adding the nodes and edges according to the SFILES~2.0 notation rules. Note that the node numbering happens before the string traversal and is according to the order of occurrence in the SFILES~2.0 string. The latter implies that the node numbers of the original flowsheet graph and the reconstructed version might differ. 
However, the topology of the translated flowsheet information is preserved. 

\section{Illustrative examples}
\begin{figure}[h!]
	\centering
	\includegraphics[width=\textwidth]{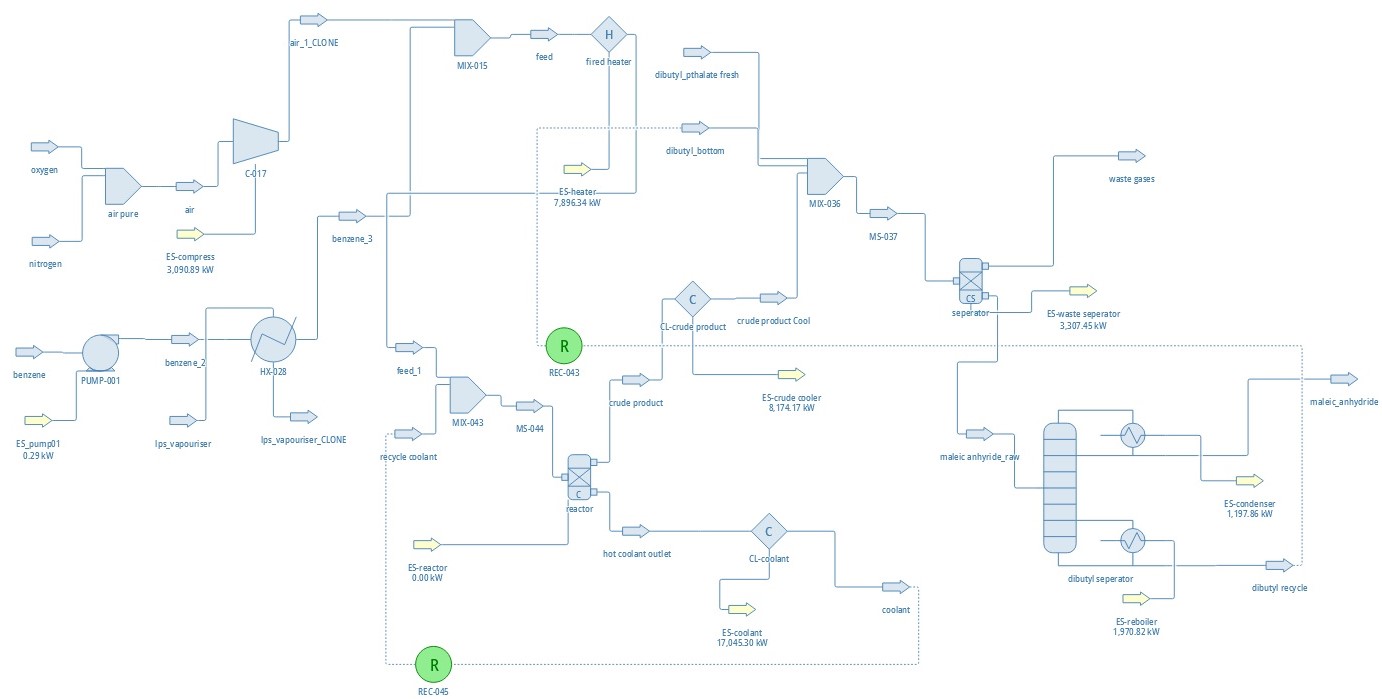}
	\caption{Process flow diagram for maleic anhydride production from benzene}
	\label{fig:PFD_maleic_anhydride}
\end{figure}

\begin{figure}[h!]
	\centering
	\includegraphics[width=\textwidth]{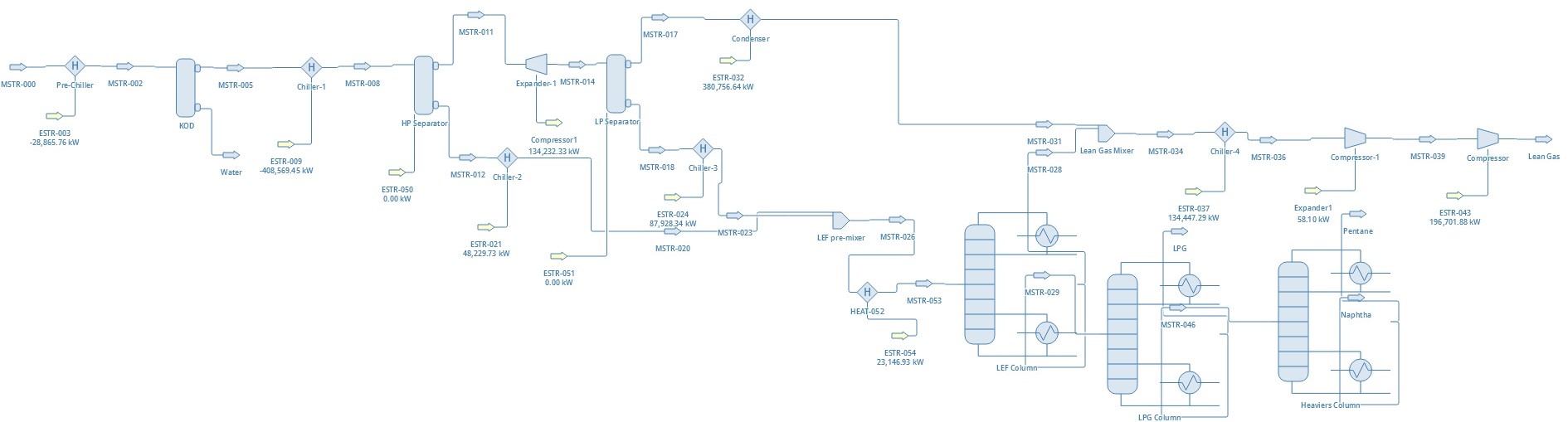}
	\caption{Process flow diagram of a natural gas processing unit}
	\label{fig:PFD_NGprocess}
\end{figure}

\begin{figure}[h!]
	\centering
	\includegraphics[width=\textwidth]{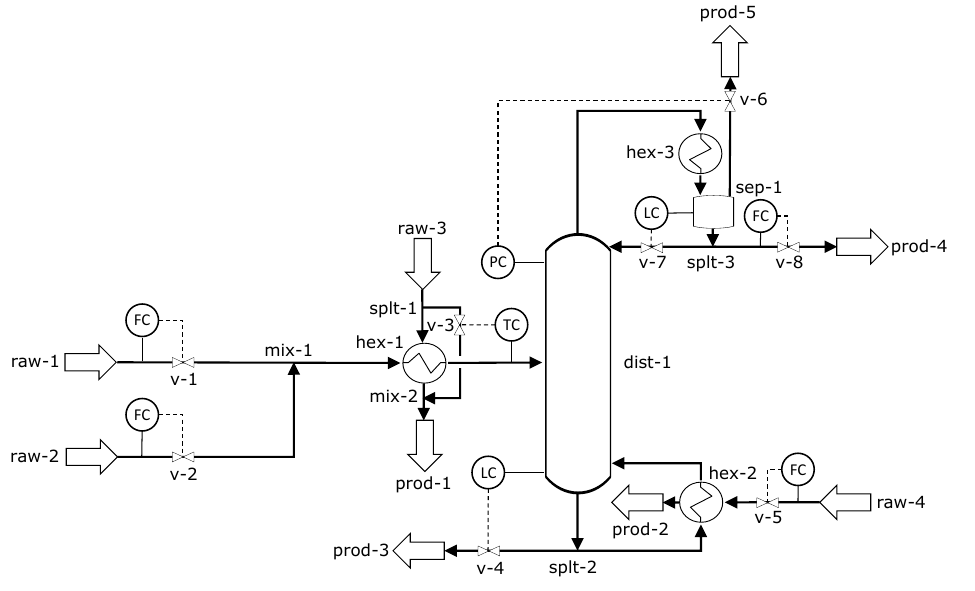}
	\caption{Process flowsheet of a distillation column with control structure}
	\label{fig:PID_appendix}
\end{figure}

\begin{figure}[h!]
	\centering
	\includegraphics[width=\textwidth]{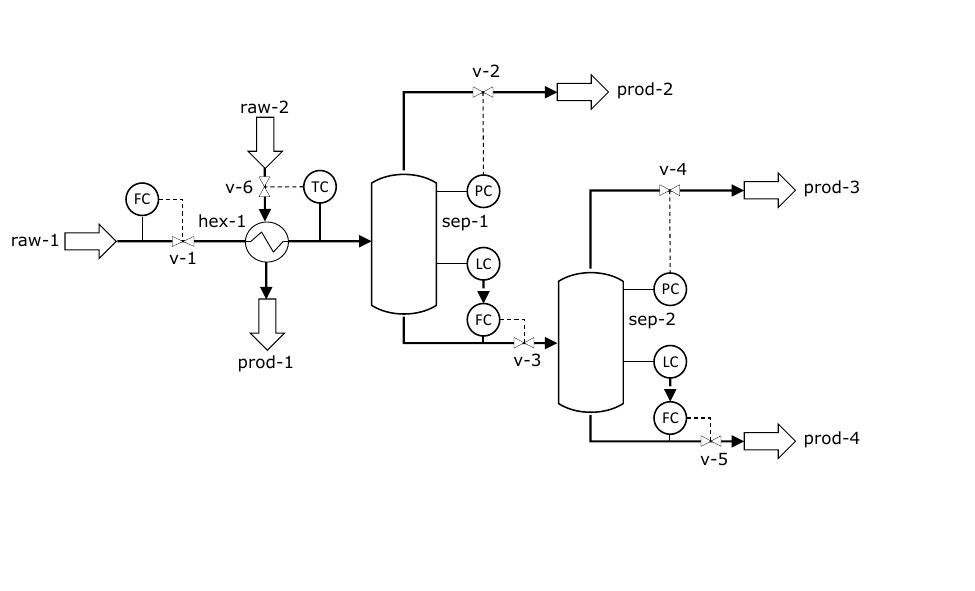}
	\caption{Two-stage flash process flowsheet with control structure}
	\label{fig:PID_appendix2}
\end{figure}

This Section provides additional examples of flowsheets with a higher number of unit operations and control structures. Figure \ref{fig:PFD_maleic_anhydride} shows the process flow diagram for the production of maleic anhydride from benzene which was extracted from a DWSIM simulation file. The corresponding flowsheet graph contains 22 nodes. Converting the flowsheet graph to the SFILES~2.0 representation yields:
\begin{lstlisting}[mathescape=true]
(raw)(pp)(hex){1}(mix)<&$|$(raw)(mix)<&$|$(raw)&$|$(comp)&$|$(hex)(mix)<1(r)[(hex)1](hex)(mix)<2<&$|$(raw)&$|$(sep)[{tout}(prod)]{bout}(dist){bout}2{tout}(prod)n$|$(raw)(hex){1}(prod).
\end{lstlisting}
Figure \ref{fig:PFD_NGprocess} shows the PFD of a natural gas processing unit with many branches. The corresponding SFILES~2.0 string is:
\begin{lstlisting}[mathescape=true]
(raw)(hex)(flash)[{bout}(prod)]{tout}(hex)(flash)[{tout}(turb)(flash)[{bout}(hex)(mix)<2(hex)(dist)[{bout}(dist)[{tout}(prod)]{bout}(dist)[{bout}(prod)]{tout}(prod)]{tout}(mix)<1(hex)(comp)(comp)(prod)]{tout}(hex)1]{bout}(hex)2
\end{lstlisting}
Figure \ref{fig:PID_appendix} shows a P\&ID of a distillation column with a high number of unit operations and control structures. The corresponding SFILES~2.0 string is:
\begin{lstlisting}[mathescape=true]
(raw)(C){FC}_1(v)<_1(mix)<&$|$(raw)(C){FC}_2(v)&<_2$|$(hex){1}(C){TC}_3(dist)<1<2[(C){PC}_4][(C){LC}_5][{tout}(hex)(sep)[(C){LC}_6][(v)<_4(prod)](splt)[(C){FC}_7(v)<_7(prod)](v)1<_6]{bout}(splt)[(v)<_5(prod)](hex){2}2n$|$(raw)(splt)[(hex){1}(mix)<3(prod)](v)3<_3n$|$(raw)(C){FC}_8(v)<_8(hex){2}(prod).    
\end{lstlisting}
Figure \ref{fig:PID_appendix2} shows the P\&ID of a two-stage flash process with control structures. The corresponding SFILES~2.0 string is:
\begin{lstlisting}[mathescape=true]
(raw)(C){FC}_1(v)<_1(hex){1}(C){TC}_2(sep)[(C){PC}_3][(C){LC}_4][(v)<_3(prod)](C){FC}_5<_4(v)<_5(sep)[(C){PC}_6][(C){LC}_7][(v)<_6(prod)](C){FC}_8<_7(v)<_8(prod)n$|$(raw)(v)<_2(hex){1}(prod).    
\end{lstlisting}

\section{Conclusions}

This paper is a proposition of the SFILES~2.0, containing modifications and extensions of the previously used SFILES. The development aims to include all essential topological information of flowsheets in the SFILES representation, such as a distinction between top- and bottom branches of unit operations. Moreover, the SFILES~2.0 includes a concept to describe control structures, which are mandatory for the operation of chemical plants. This extends the applicability of SFILES~2.0 from PFDs to P\&IDs, which are the predominant diagram types utilized during the development and operation of chemical plants. To leverage the full potential regarding future databases, the SFILES~2.0 notation comes with naming conventions for the unit operations and a set of standardized stream tags. Eventually, the implementation of the reversible conversion between flowsheet graph and SFILES~2.0 strings is openly accessible to enable researchers and engineers to write or read SFILES~2.0 strings. This work attempts to lay the foundation for creating an SFILES~2.0-based database for PFDs and P\&IDs, ideally containing a large variety of chemical processes. 

\section{Acknowledgements}
This publication is part of the project “ChemEng KG – The Chemical Engineering Knowledge Graph” with project number 203.001.107 of the research programme “Open Science (OS) Fund 2020/2021” which is (partly) financed by the Dutch Research Council (NWO).




\end{document}